\providecommand{\pgfsyspdfmark}[3]{}

\documentclass[PAPER, atlasdraft=false, cernpreprint, USenglish, texmf, orcidlogo]{atlasdoc}
\usepackage{atlaspackage}
\usepackage{atlasbiblatex}

\usepackage{atlasphysics}
\graphicspath{{logos/}{figures/}}

\usepackage{ANA-STDM-2020-17-PAPER-defs}
\usepackage{listings}
\usepackage{xcolor}
\usepackage{tikz}

\definecolor{codegreen}{rgb}{0,0.6,0}
\definecolor{codegray}{rgb}{0.5,0.5,0.5}
\definecolor{codepurple}{rgb}{0.58,0,0.82}
\definecolor{backcolour}{rgb}{0.95,0.95,0.92}

\lstdefinestyle{mystyle}{
backgroundcolor=\color{backcolour}, commentstyle=\color{codegreen},
keywordstyle=\color{magenta},
numberstyle=\tiny\color{codegray},
stringstyle=\color{codepurple},
basicstyle=\ttfamily\footnotesize,
breakatwhitespace=false,
breaklines=true,
captionpos=b,
keepspaces=true,
numbers=left,
numbersep=5pt,
showspaces=false,
showstringspaces=false,
showtabs=false,
tabsize=2
}

\AtlasTitle{A simultaneous unbinned differential cross section measurement of twenty-four $Z$+jets kinematic observables with the ATLAS detector}

\AtlasAbstract{%

$Z$ boson events at the Large Hadron Collider can be selected with high purity and are sensitive to a diverse range of QCD phenomena.
As a result, these events are often used to probe the nature of the strong force, improve Monte Carlo event generators, and search for deviations from Standard Model predictions.
All previous measurements of $Z$ boson production characterize the event properties using a small number of observables and present the results as differential cross sections in predetermined bins. In this analysis, a machine learning method called \textsc{OmniFold} is used to produce a simultaneous measurement of twenty-four $Z$+jets observables using $139$ fb$^{-1}$ of proton-proton collisions at $\sqrt{s}=13$ TeV collected with the ATLAS detector. Unlike any previous fiducial differential cross-section measurement, this result is presented unbinned as a dataset of particle-level events, allowing for flexible re-use in a variety of contexts and for new observables to be constructed from the twenty-four measured observables.

}

\AtlasRefCode{STDM-2020-17}

\PreprintIdNumber{CERN-EP-2024-132}

\AtlasJournal{Phys.\ Rev.\ Lett.}
\AtlasJournalRef{Phys.Rev.Lett. 133 (2024) 26, 261803}
\AtlasDOI{10.1103/PhysRevLett.133.261803}

\hypersetup{pdftitle={ATLAS document},pdfauthor={The ATLAS Collaboration}}

\begin{document}

\maketitle

% The next lines are included from the .//multifold-paper-body.tex input file
The production of $Z$~bosons is a standard candle process at the Large Hadron Collider~(LHC), used for various purposes, such as precision tests of the Standard Model, detector calibration, and testing new analysis methods. The large $pp\to Z+X$ cross section, in combination with the easily identifiable $Z\to\ell\ell$ decay (with $\ell \in \{e,\mu\}$), makes it possible to collect event samples with high purity and efficiency.
When the $Z$ boson is produced at large transverse momentum, it is usually accompanied by an associated hadronic recoil that is collimated into one or more jets.
Measurements of $Z$+jets production are crucial for many purposes,
including powerful tests of perturbative quantum chromodynamics (QCD)~\cite{Drell:1970wh,Hamberg:1990np,Melnikov:2006kv,Gross:1973ju,Politzer:1974fr} and improvements of the parameters used in parton shower Monte Carlo calculations~\cite{Larkoski:2017jix,Kogler:2018hem} and tunes of the underlying event~\cite{ATL-PHYS-PUB-2014-021, STDM-2012-23}.

Numerous measurements of $Z$+jets production that probe the kinematic properties of the $Z$~boson and the associated jets were performed at the Tevatron~\cite{CDF:2014hdx,D0:2009rnw} and at the LHC~\cite{STDM-2012-04,STDM-2016-01,STDM-2018-49,STDM-2017-27,CMS-SMP-12-017,CMS-SMP-14-013,CMS-SMP-19-009,CMS-SMP-16-018, CMS-SMP-21-003}, including several dedicated measurements of the internal structure of the associated jets~\cite{STDM-2015-14,CMS-SMP-12-019,CMS-SMP-20-010,LHCb:2019qoc}.
Each of these measurements takes the form of a binned fiducial differential cross section at the particle level by fully correcting for detector effects using \textit{unfolding} methods~\cite{Cowan:2002in,Blobel:jla,Brenner:2019lmf}.

The most widely used unfolding methods employ forms of regularized matrix inversion~\cite{DAgostini:1994fjx,Hocker:1995kb,Schmitt:2012kp}.
This analysis presents developments that address four features of traditional unfolding techniques that potentially limit the future utility of the published data.
First, the target observables must be specified prior to unfolding and cannot be changed after the measurement.  Second, the binning of the observables must be fixed at the start of the measurement.  Third, due to the binned nature of existing techniques, most measurements are done as a function of a single observable, and only occasionally in bins of two or three observables (e.g.\ Refs.~\cite{STDM-2016-03,STDM-2016-04,CMS-SMP-16-010,CMS-SMP-16-011}).
Finally, existing methods can have large uncertainties associated with biases in the detector response due to mismodelling by the Monte Carlo simulation of observables other than the ones directly measured. %

Recently proposed machine learning methods address these challenges directly~\cite{Glazov:2017vni,Arratia:2021otl}.  Such methods use discriminative~\cite{Glazov:2017vni,bunse2018unification,Ruhe2019MiningFS,Andreassen:2019cjw,Andreassen:2021zzk,Arratia:2022wny,Chan:2023tbf} or generative~\cite{Datta:2018mwd,Bellagente:2019uyp,Bellagente:2020piv,Vandegar:2020yvw,Howard:2021pos,Backes:2022vmn,Shmakov:2023kjj,Alghamdi:2023emm,Diefenbacher:2023wec} neural networks (NNs) to readily process dozens of input observables in an unbinned manner. One discriminative approach is \omnifold~\cite{Andreassen:2019cjw,Andreassen:2021zzk}, an iterative method that generalizes to unbinned data the widely used Lucy-Richardson deconvolution approach~\cite{Lucy:1974yx,Richardson:72} (also known as Iterative Bayesian unfolding or IBU~\cite{DAgostini:1994fjx}). This method has recently been applied to perform the first unbinned studies\footnote{While the unfolding is unbinned, the resulting cross sections in these studies are all released publicly in a binned format.} of hadronic final states with data from H1~\cite{H1:2021wkz,H1:2023fzk}, LHCb~\cite{LHCb:2022rky}, CMS~\cite{Komiske:2022vxg}, and STAR~\cite{Song:2023sxb}.  \omnifold{} learns a correction (assigned as event weights) to an initial set of simulated events instead of the more difficult task of learning to produce new events, as is done in generative approaches.
As these methods are multidimensional, they can better account for variable dependency of the detector response that improves the measurement precision.

This Letter presents unbinned differential cross sections for $Z$+jets events in the dimuon channel $Z/\gamma^\ast\to\mu\mu$ using the \omnifold{} method.
The result constitutes a precision measurement in its own right, with multiple novel use cases as described below, and also serves as a proof-of-principle application of the \omnifold{} method to provide an unbinned, high-dimensional measurement with full covariance for public use.
The analysis is performed using the full Run~2 proton--proton dataset collected by the ATLAS detector~\cite{PERF-2007-01} at $\sqrt{s}=\SI{13}{\TeV}$, corresponding to an integrated luminosity of $\mathcal{L}=\SI{139}{\per\femto\barn}$.
The measurements are at the particle level, defined by final-state stable particles with mean lifetime satisfying $c\tau > \SI{10}{\milli\meter}$.
The fiducial volume requires two muons not originating from the decay of hadrons that each satisfy transverse momenta $\pt>\SI{25}{\GeV}$ and $|\eta|<2.5$.\footnote{ATLAS uses a right-handed coordinate system with its origin at the nominal interaction point (IP) in the center of the detector and the \(z\)-axis along the beam pipe. The \(x\)-axis points from the IP to the center of the LHC ring, and the \(y\)-axis points upward. Cylindrical coordinates $(r, \phi)$ are used in the transverse plane, $\phi$ being the azimuthal angle around the beam pipe. The pseudorapidity is defined in terms of the polar angle $\theta$ as $\eta \equiv - \ln\tan(\theta/2)$ and is equal to the rapidity $y \equiv 0.5\,\ln\left((E+p_\mathrm{z})/(E-p_\mathrm{z})\right)$ in the relativistic limit. Angular distance is measured in units of $\Delta R = \sqrt{(\Delta y)^2+(\Delta \phi)^2}$.}
The final-state muon is ``dressed'', such that collinear radiation of photons within a cone of $\Delta R = 0.1$ are added to its four momentum.
The muons are further required to have opposite charges, dimuon invariant mass $m_{\mu\mu}\in(81,101)$\,\GeV, and $\ptll>\SI{200}{\GeV}$.
The last criterion selects an unbiased sample of high-\pt{} jets, allowing jet properties to be probed in a previously underexplored kinematic regime, and reduces the size of the dataset, which simplifies the computational challenge for the unfolding method.
The jets are reconstructed from charged particles with $\pt > \SI{0.5}{\GeV}$ and $|\eta|<2.5$ using the anti-$k_t$ algorithm~\cite{Cacciari:2008gp,Fastjet} with $R = 0.4$; charged particles corresponding to prompt leptons, such as muons from $Z\to\mu\mu$, are excluded from the jet finding. Charged particles are used due to the precision with which they can be measured and the availability of full uncertainties. No additional acceptance requirements are placed on the jets, as these are implicitly set by the charged particle selection criteria.
The 24 measured event observables are:
\begin{itemize}
\item
\ptll{} and \yll{} of the dimuon system that probe the $Z$~boson production kinematics;
\item
the kinematics of the two muons defined by \ptlm, \ptsm, \etalm, \etasm, \philm, \phism{}, which probe the $Z$~boson decay kinematics;
\item
the kinematics of the two leading charged particle jets defined by \ptlj, \ptsj, \ylj{}, \ysj, \philj, and \phisj{}; and
\item
the masses (\mlj, \msj), charged particle multiplicities (\Nclj, \Ncsj) and $N$-subjettiness quantities \taulj{1},
\tausj{1}, \taulj{2}, \tausj{2}, \taulj{3} and \tausj{3}~\cite{Thaler:2010tr,Thaler:2011gf} that probe the substructure of the two leading charged particle jets.
\end{itemize}
There is a significant overlap in observables between the \omnifold{} analysis and the ones used to produce the ATLAS A14 parameter set (tune)~\cite{ATL-PHYS-PUB-2014-021} of the \PYTHIA{} event generator~\cite{Sjostrand:2007gs,Sjostrand:2014zea}.
A natural application of this measurement would hence be to create precise event generator tunes improving the modeling of the parton showers, hadronization and the underlying event.
Other uses could include studies of jet substructure and jet flavor; for example, selecting jets back-to-back with the $Z$~boson should yield quark-like jets, while wide angle radiation would give more gluon-like jets.
Since the measurement is unbinned and probes a wide \pT{}-range of jets, it is straightforward to switch between observables and study various quantities as a function of other quantities (e.g.\ jet $m$ vs \pt{}, \Nclj{} vs \ylj{}, etc.).

The ATLAS detector has forward--backward symmetric cylindrical geometry and a near \(4\pi\) coverage in solid angle.
It includes an inner detector (ID) for charged particle tracking covering \(|\eta| < 2.5\) surrounded by a thin
solenoid providing an axial field of \SI{2}{\tesla},
electromagnetic and hadron calorimeters, and a muon spectrometer (MS).
A two-level trigger system is used to select events~\cite{TRIG-2016-01}.
An extensive software suite~\cite{SOFT-2022-02} is used for all aspects of data collection, curation, and analysis.

Data events are collected using single-muon triggers~\cite{TRIG-2018-01}.
Muons are reconstructed by matching charged particle tracks in the ID and MS, accounting for energy loss in the calorimeters~\cite{MUON-2018-03}. They are required to fulfill \textit{Medium} identification criteria and \textit{PflowLoose} isolation~\cite{MUON-2018-03}, and must satisfy $|d_{0}|/\sigma_{d_{0}} < 3$ and $|z_{0}\sin(\theta)| < \SI{0.5}{\milli\meter}$ for their transverse and longitudinal distances to the point of closest approach to the beamspot, ensuring that they originate from the interaction vertex, i.e.\ the primary vertex with the highest sum of associated track $p_\mathrm{T}^2$.
Charged particle tracks used for jet building are required to fulfill \textit{Loose} quality criteria, a \textit{Tight} track-to-vertex matching criterion~\cite{ATL-PHYS-PUB-2015-051}, and must not be used by the selected muons. Identical kinematic requirements are applied to the reconstructed quantities as those used to define the particle-level fiducial volume.

Monte Carlo (MC) simulated samples are used to provide theoretical predictions both at the reconstructed and particle level, and to perform the unfolding. Most $Z$+jets events that satisfy the selection originate from the Drell--Yan process.
Two MC samples are used to model this process: a nominal prediction is provided by \MGNLO{2.6.5}~\cite{Alwall:2014hca,Frederix:2012ps,PMGR-2021-01} interfaced to \PYTHIA[8.240] (denoted \MADGRAPH{} in the following), and \SHERPA[2.2.11]~\cite{Bothmann:2019yzt,PMGR-2021-01} provides an alternative.
Both samples use the \NNPDF[3.0nnlo] parton distribution functions (PDF) set~\cite{Ball:2014uwa}; additional details for these samples are given in Ref.~\cite{PMGR-2021-01}.
Contributions from electroweak production of $Z$+jets are provided using \HERWIG[7.2]~\cite{Bahr:2008pv,Bellm:2019zci} interfaced with \VBFNLO[v3.0.0]~\cite{vbfnlo} using the MMHT2014LO PDF set~\cite{Harland-Lang:2014zoa}. Diboson $ZV$$\to$$Zjj$ production is modeled at NLO accuracy by \SHERPA[2.2.1] or 2.2.2~\cite{Bothmann:2019yzt} using \NNPDF[3.0nnlo].
Background contributions from top processes ($t\bar{t}$, $tV$, single top) are modeled by \POWHEG[v2]~\cite{Alioli:2010xd} interfaced to \PYTHIA[8.230].
The detector response is simulated using a dedicated \GEANT{}-based model~\cite{Agostinelli:2002hh} of the ATLAS detector~\cite{SOFT-2010-01}. Simulated inelastic minimum-bias events are overlaid to model additional $pp$ collisions in the same and neighboring bunch crossings (``pileup'')~\cite{SOFT-2010-01}. Simulated events are reconstructed using the same procedure as for data.

After the event selection, a pure $Z$+jets signal sample is obtained with a composition of about 95\% Drell--Yan, 3\% diboson (primarily $ZV\to\mu\mu jj$), and 2\% electroweak $Zjj$. The fraction of non-Drell--Yan $Z$+jets increases with \ptll{} and reaches about 10\% at $\ptll > \SI{500}{\GeV}$. The fraction of diboson events is also sensitive to $\mlj$, and reaches $\sim 8\%$ for $\mlj > \SI{45}{\GeV}$.
The analysis measures inclusive $Z$+jets production and makes no attempt to separate these processes. Only the Drell--Yan production is used in the simulation for the unfolding itself, and the full difference between this and the result in which all $Z$+jets are included
is found to be small and taken as a conservative estimate of the process composition uncertainty.
Backgrounds of about 0.2\% arise due to the top processes (mostly $t\bar{t}\to\mu\mu\nu\nu jj$ and $tW\to\mu\mu\nu jj$).  Contributions from events without two muons from the hard-scatter are found to be negligible. The top background increases in regions with significant jet activity and is estimated to be $\sim 2\%$ for both $\ptlj>\SI{300}{\GeV}$ and $\Nclj>25$. As the background is small, it is not subtracted in the measurement; an estimate of its contribution is instead assigned as an uncertainty.

The \omnifold{}-based unfolding produces event weights that are applied to the \MADGRAPH{} Drell--Yan $Z$+jets sample at particle level (see the Appendix).  The number of iterations was fixed to five following a dedicated study that found unfolding performance plateaued around that number.
The analysis is performed in a phase-space slightly larger ($\ptll>\SI{190}{\GeV}$) than the fiducial volume ($\ptll>\SI{200}{\GeV}$) in order to reduce migration uncertainties,
which also includes the final normalization of the results such that it provides fiducial cross sections, $\sigma_\mathrm{fid}$.
During this step, all \omnifold{} weights are scaled by a constant to fulfill
the relation $\mathcal{L}\,\sigma_\mathrm{fid}\,\varepsilon/f_\mathrm{fid} = n_\mathrm{data}$, where the overall efficiency $\varepsilon$ and fiducial factor $f_\mathrm{fid}$ is evaluated using the same MC as used for unfolding, and $n_\mathrm{data}$ is the data count.
After normalization, the sum of weights in any subset of the \omnifold{} dataset can be interpreted as an estimate of its associated cross section. The final output of \omnifold{} is the original particle-level simulated event sample with additional event weights that can be used to define measurements of the cross-section of fiducial subregions (bins) defined by the 24 observables subject to the precision of the measurement. This includes differential cross sections of any of the 24 input variables or any combinations of those observables.

All NNs are constructed in TensorFlow~\cite{Abadi:2016kxi,Abadi:2016kic}, with three hidden layers of 200 nodes each with Rectified Linear Unit (ReLU)~\cite{relu} activation functions and a sigmoid final activation function.
Two main challenges were faced regarding obtaining reliable and accurate NN outputs.
The first is regarding MC event weights of the input samples, which initially had a large spread and often were negative that causes issues for the performance.
This is addressed by pre-processing the MC sample such that negative weights are removed and the spread of weights is reduced as described in Ref.~\cite{Nachman:2020fff}.
The second challenge is intrinsic to the NNs themselves, as the classifier output can vary slightly due to the randomly-initialized starting weights.
To stabilize the result, an ensemble of 100 NNs are created for each training, and the weight is taken as the median of the 100 NN weights.

Uncertainties on the unfolded result are evaluated using error propagation.
Perturbations are introduced to the input samples by an amount commensurate with the uncertainty variation in question, and the full analysis (unfolding and normalization) is repeated, resulting in \omnifold{} weights that differ from the nominal weights.
The measurement central value is obtained with the nominal weight, and a total of 250 variation weights, each used to estimate the uncertainty.

Systematic uncertainties are split into 25 components that are each treated as independent.
Experimental sources of uncertainty include systematic bias due to:
the muon efficiency and calibration~\cite{MUON-2018-03}; track reconstruction~\cite{ATL-PHYS-PUB-2015-051}; pileup modeling; and, the luminosity measurement~\cite{ATLAS-CONF-2019-021}. Theoretical uncertainties are evaluated for variations of PDF and $\alpha_s$ choices~\cite{PMGR-2021-01}, QCD scales~\cite{PMGR-2021-01}, and the generator tune~\cite{STDM-2012-23}.
Imperfections inherent to the choice of Monte Carlo simulation contribute to two sources of systematic uncertainty: the sensitivity to differences in the underlying truth distribution of the measured observables, and the sensitivity to mismodelling of detector effects by the Monte Carlo generator that are not captured by the chosen input variables.
The uncertainty (``unfolding prior'') for the imperfect particle-level shape of the initial MC sample is assessed by reweighting the nominal MC sample at particle level such that it approximately agrees with data for the 24 observables. This reweighting function is constructed using a sequence of one-dimensional Gaussian-kernel functions, iteratively obtained from the data-to-MC ratio.
The obtained data-driven correction is applied to the MC to obtain an ``Asimov dataset'' used as the input to the measurement, and the difference between the resulting measurement and the corresponded reweighted particle level MC is taken as the uncertainty.
An uncertainty in the top-quark background is assessed as the full difference between measurements performed using two Asimov datasets constructed from MC predictions with and without the top-quark contribution.
To assess the dependence on the detector response from modeling of features not included in the unfolding (``hidden variable uncertainty''), the measurement is performed with the alternative Drell--Yan MC sample with the particle-level shape of the 24 observables adjusted to match the nominal MC sample (see also \cite{CMS-PAS-SMP-23-008, ATLAS-CONF-2023-028, ATLAS-CONF-2023-019} for similar procedures).
Similarly, modeling of the non-Drell--Yan components (EW $Zjj$ and $ZV$) are assessed as the full difference by performing the measurement with and without these components added. Both of these variations are provided as separate event datasets that provided varied measurements (two-point uncertainties).

Four types of stochastic uncertainties are assessed:
statistical uncertainties on the data and the MC training sets are each assessed by bootstrapping~\cite{ATL-PHYS-PUB-2021-011} (100 and 25 weights, respectively); an uncertainty due to the NN stability is calculated from the standard error on the median of weights of the 100 individual NNs; and an additional uncertainty is assigned for the limited statistics of the nominal event dataset.
Overall, the total uncertainty in most bins chosen to illustrate the final measurement is between 3\% and 5\%, but can grow as large as 15\% in tails of distributions.
The unfolding uncertainty from the unfolding prior and hidden variables tends to be the dominant contributor for many observables, in particular for the ten jet substructure variables (see Figure~\ref{fig:nom}(d)).

\usetikzlibrary{calc}

\begin{figure} %
\begin{tikzpicture}[remember picture]
\node[anchor=south west, inner sep=0] at (0,0) {
\includegraphics[width=0.245\textwidth]{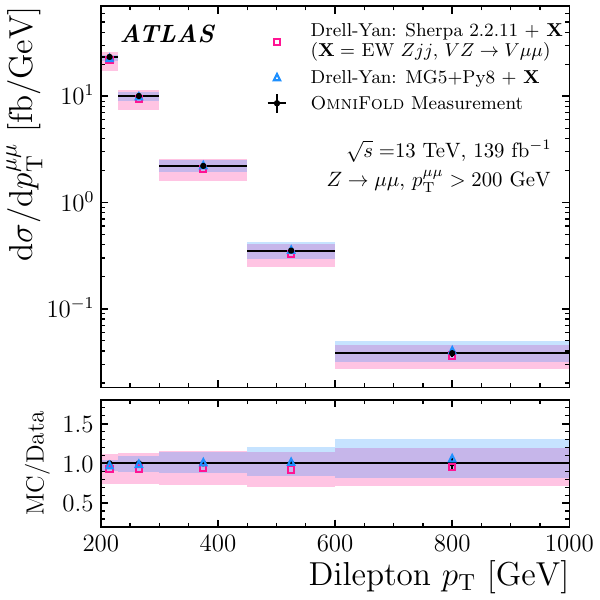}
\includegraphics[width=0.245\textwidth]{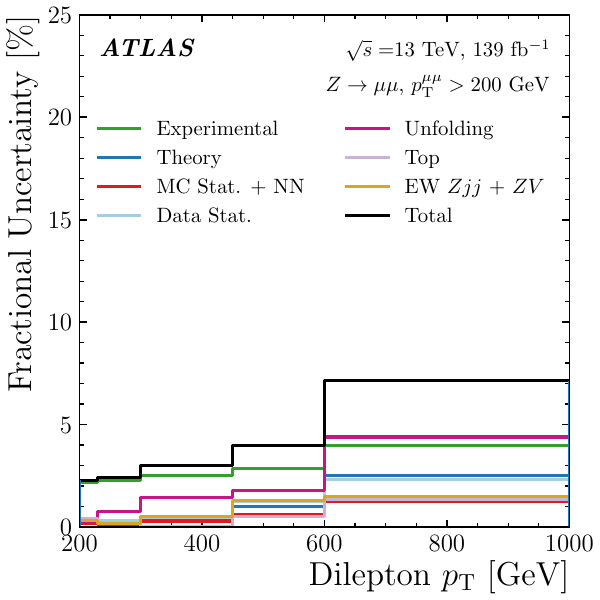}
\includegraphics[width=0.245\textwidth]{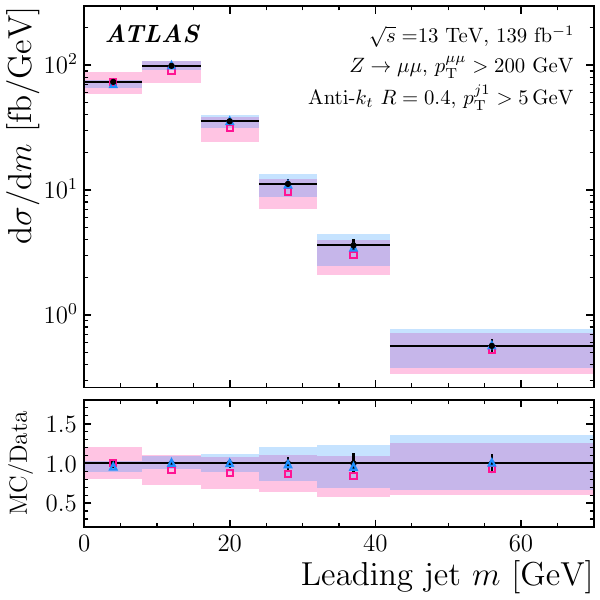}
\includegraphics[width=0.245\textwidth]{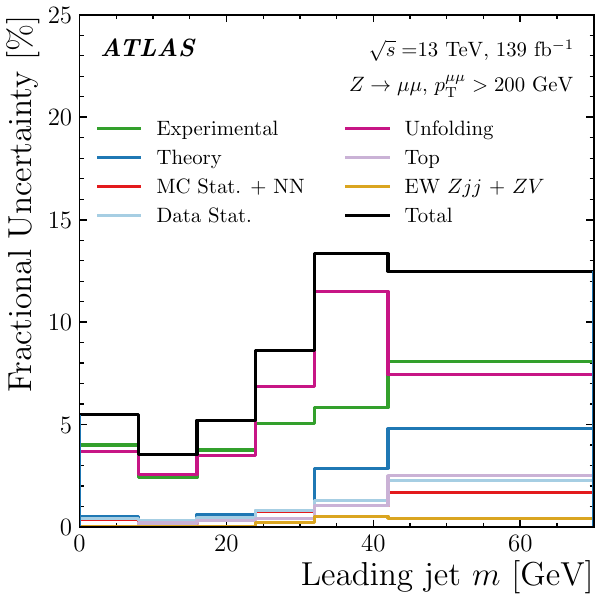}
};
\coordinate (a) at ($(current bounding box.south west) + (10.2mm, 17mm)$);
\node[align=center, font=\small] at (a) {(a)};
\node[align=center, font=\small] at ([xshift=39.5mm]a) {(b)};
\node[align=center, font=\small] at ([xshift=81mm]a) {(c)};
\node[align=center, font=\small] at ([xshift=120.5mm]a){(d)};
\end{tikzpicture}
\caption{
Measured differential cross sections compared with particle-level predictions from \SHERPA{} and \MADGRAPH{} for two of the 24 directly measured observables:
(a)~\ptll{} with its (b)~associated uncertainty breakdown; and, (c)~\mlj{} with its (d)~associated uncertainty breakdown.
For display purposes, binned (marginal) distributions are shown, though the measurement itself is unbinned and 24-dimensional.
}
\label{fig:nom}
\end{figure}

The measured differential cross sections of \ptll{} and \mlj, in comparison with two MC predictions, are shown in Figure~\ref{fig:nom} along with breakdowns of the associated measurement uncertainties. Plots of the measured spectra, associated uncertainty breakdown and correlation matrices for all of the 24 directly measured observables are provided as Supplemental Material~\cite{supplemental}. The total fiducial measured cross section is 1,808$\pm$42\,fb. The differential measurements are significantly more precise than the predictions, in particular with respect to \SHERPA.
\MADGRAPH{} generally models the data better than \SHERPA, except for \tausj{1}, \tausj{2} and \tausj{3}.
The measurement is publicly available as event datasets that contain the 24 observables and a series of event weights that define the measurement and systematic uncertainties via Refs.~\cite{public_multifold, zenodo}.

\begin{figure} %
\begin{tikzpicture}[remember picture]
\node[anchor=south west, inner sep=0] at (0,0) {
\includegraphics[width=0.245\textwidth]{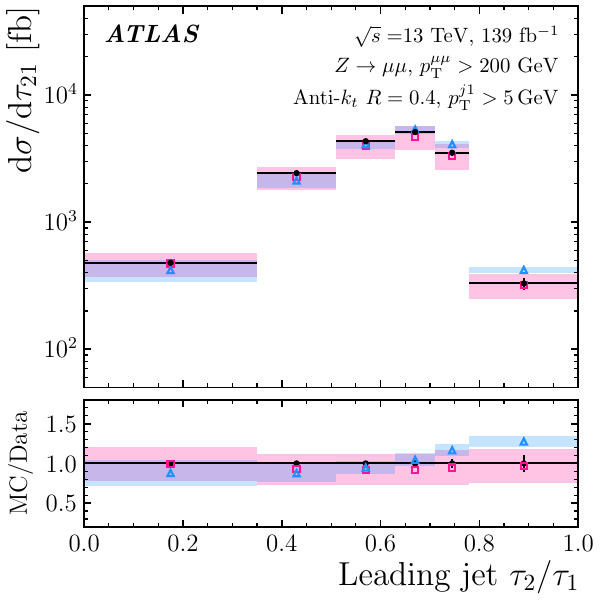}
\includegraphics[width=0.245\textwidth]{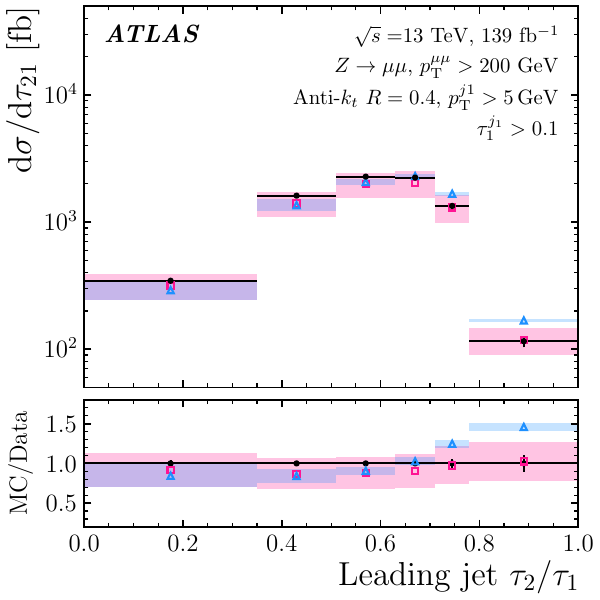}
\includegraphics[width=0.245\textwidth]{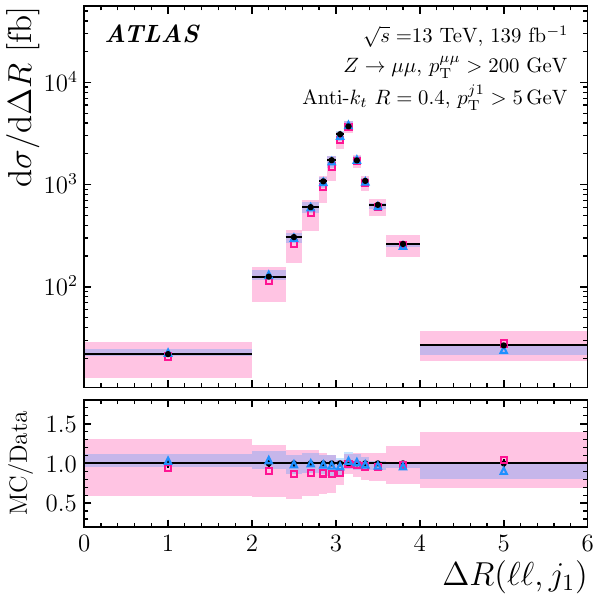}
\includegraphics[width=0.25\textwidth]{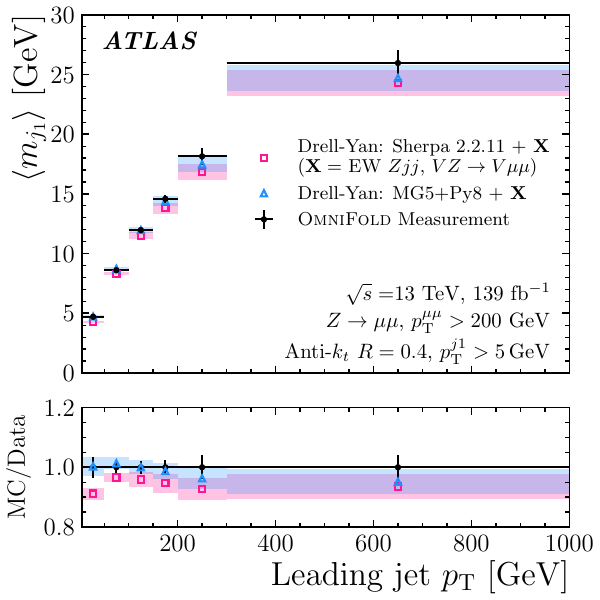}
};
\coordinate (e) at ($(current bounding box.south west) + (9mm, 17mm)$);
\node[align=center, font=\small] at (e) {(a)};
\node[align=center, font=\small] at ([xshift=40.5mm]e) {(b)};
\node[align=center, font=\small] at ([xshift=92mm]e) {(c)};
\node[align=center, font=\small] at ([xshift=122mm,yshift=1mm]e){(d)};
\end{tikzpicture}
\caption{
Four measurements of quantities constructed from several of the 24 input observables, along with particle-level predictions from \SHERPA{} and \MADGRAPH{}: the jet substructure observable $\tau^{j1}_{21} = \taulj{2}/\taulj{1}$ in (a)~the inclusive region, and (b)~the reduced region defined by $\taulj{1} > 0.1$; (c)~$\Delta R$ between the dilepton system and the leading jet; and, (d)~the average \mlj{} as a function of \ptlj{}.
}
\label{fig:derived}
\end{figure}

Figure~\ref{fig:derived} presents additional results constructed from the nominal measurement that highlight its flexibility of use.
Figures~\ref{fig:derived}(a)--(c) show the differential cross sections of ``derived'' variables that were not directly unfolded, namely $\tau_{21} = \tau_2/\tau_1$ (the most widely used observable for hadronic $W/Z$ boson identification~\cite{Thaler:2011gf,Thaler:2010tr}) and $\Delta R(\ell\ell, j1)$ (sensitive to higher-order effects). These observables are functions of two and eight of the 24 input variables, respectively. In the inclusive region, $\tau_{21}$ is not infrared and collinear (IRC) safe and therefore has no fixed-order perturbative expansion in $\alpha_S$.
It has been shown~\cite{Larkoski:2013paa} that $\tau_{21}$ becomes IRC safe when applying a requirement on $\tau_1$, and Figure~\ref{fig:derived}(b) shows an unprecedented measurement of $\tau_{21}$ in an IRC-safe fiducial volume defined by $\tau_1^{j_1} > 0.1$.
Figure~\ref{fig:derived}(d) shows a measurement of the average \mlj{} in bins of \ptlj{}, providing an example of a distribution useful for MC tuning.
All of the derived observables can be calculated from the \omnifold{} data event-by-event and be used to define bins with associated cross sections and uncertainties, just as for any of the 24 input variables.

The analysis was validated using a ``pseudo-data'' sample constructed by reweighting the particle-level quantities in the alternative MC sample to resemble the real reconstruction-level data. Two such samples were generated: a high-statistics sample with weighted events and a data-like sample with unit weights obtained by bootstrapping the former sample. The full analysis was performed on the unit-weight pseudo-dataset, and the unfolding bias uncertainty was evaluated by assessing the closure between the \omnifold{} result with the particle-level target. Chi-squared compatibility tests between the obtained measurements and the known underlying particle-level distributions for each of the 24+2 observables all yielded $p$-values $>0.07$, except for $p_T^{j_1}$, with a $p$-value of 0.038.
A full analysis of the pseudo-dataset
was also performed using IBU, where each observable was measured individually with the same binning and input data.
The central values of the IBU measurements agree well within precision with the \omnifold{} result. The IBU experimental, theoretical, unfolding and statistical uncertainty components are also similar to those of the \omnifold{} measurement in magnitude, shape, and resulting covariance. IBU also demonstrates a very similar performance in the closure test against the target.
The total uncertainty was found overall to be similar, but tends to be somewhat larger for the \omnifold{} measurement, primarily due to the NN initialization uncertainty that does not apply to IBU.
The average bin uncertainty across all 24 observables was found to be 3.0\% for IBU and 3.9\% for \omnifold{}.
The uncertainty due to hidden variables does decrease for certain variables, but not generally,
which is likely an indication that the detector response is not strongly covariate with the variables used for this measurement.

The \omnifold\ result was then validated by performing $\chi^2$ tests of differential spectra in dedicated kinematic subregions: high \ptll{} ($\ptll > 250$\,\GeV), electroweak-enhanced ($m_{jj}>200$\,\GeV, $\Delta y_{jj}>2$), and diboson-enhanced ($\mlj>32$\,\GeV). Chi-squared tests were performed against the pseudo-data target within each subregion for all measured and several derived observables. All results yielded $p$-values greater than $0.05$, except for one observable (\mlj{}) in the electroweak-enhanced region, which had a $p$-value of 0.02. The result was also validated in two-dimensional kinematic subregions, e.g.\ for \ptll{} vs. \yll, with $p$-values $>0.05$. Ablation studies on the input variables were performed to understand the effect of removing one or two variables from the unfolding procedure. These tests indicated that apart from a small number of cases in which critical variables including \mlj, \msj, \ptlj, and \ptsj were removed, removing almost any of the other twenty-four input variables still yielded excellent agreement within the quoted unfolding uncertainties. Stress tests were also performed to ensure the result is robust to non-trivial distortions. These included randomly and deterministically shifting and stretching the input spectra. No significant bias in the final result was observed.

The results of the validation tests were used to define a set of recommendations on how to use the provided datasets based on phase space coverage in data and simulation. The \omnifold{} results are entirely unbinned, so the chosen binning is for presentation purposes only and is configurable. When choosing bins, certain best practices are recommended to help ensure that the number of MC and data events per bin yield sufficient support for the unfolding and stable uncertainties. These recommendations are detailed in the User Guide found in the Supplemental Material~\cite{supplemental}, and examples of use are provided in the interactive Python notebooks associated with the published unbinned datasets~\cite{public_multifold}.

In conclusion, this Letter presents an unbinned unfolded cross-section measurement of $Z$+jets events
using \SI{139}{\per\femto\barn} of proton--proton collisions at $\sqrt{s}=\SI{13}{\TeV}$ collected with the ATLAS detector at the LHC.
The 24 observables treated in this analysis are simultaneously unfolded using the machine learning method \omnifold. These results demonstrate that collider data can be unfolded in an unbinned manner and that the result can be re-analyzed at the event level, allowing researchers significant increased utility such as adjusting binning and constructing new observables from the 24 provided ones.
This flexibility makes it possible to probe kinematic regimes and observables not originally foreseen, which can enable numerous physics use cases including strong tests of QCD and detailed tuning of MC event generators.

% End of text imported from the .//multifold-paper-body.tex input file

%

%
%

%
%

%
\section*{Acknowledgements}
%

% The next lines are included from the .//acknowledgements/Acknowledgements.tex input file
%
%

%
%

We thank CERN for the very successful operation of the LHC and its injectors, as well as the support staff at
CERN and at our institutions worldwide without whom ATLAS could not be operated efficiently.

The crucial computing support from all WLCG partners is acknowledged gratefully, in particular from CERN, the ATLAS Tier-1 facilities at TRIUMF/SFU (Canada), NDGF (Denmark, Norway, Sweden), CC-IN2P3 (France), KIT/GridKA (Germany), INFN-CNAF (Italy), NL-T1 (Netherlands), PIC (Spain), RAL (UK) and BNL (USA), the Tier-2 facilities worldwide and large non-WLCG resource providers. Major contributors of computing resources are listed in Ref.~\cite{ATL-SOFT-PUB-2023-001}.

We gratefully acknowledge the support of ANPCyT, Argentina; YerPhI, Armenia; ARC, Australia; BMWFW and FWF, Austria; ANAS, Azerbaijan; CNPq and FAPESP, Brazil; NSERC, NRC and CFI, Canada; CERN; ANID, Chile; CAS, MOST and NSFC, China; Minciencias, Colombia; MEYS CR, Czech Republic; DNRF and DNSRC, Denmark; IN2P3-CNRS and CEA-DRF/IRFU, France; SRNSFG, Georgia; BMBF, HGF and MPG, Germany; GSRI, Greece; RGC and Hong Kong SAR, China; ISF and Benoziyo Center, Israel; INFN, Italy; MEXT and JSPS, Japan; CNRST, Morocco; NWO, Netherlands; RCN, Norway; MNiSW, Poland; FCT, Portugal; MNE/IFA, Romania; MESTD, Serbia; MSSR, Slovakia; ARIS and MVZI, Slovenia; DSI/NRF, South Africa; MICIU/AEI, Spain; SRC and Wallenberg Foundation, Sweden; SERI, SNSF and Cantons of Bern and Geneva, Switzerland; NSTC, Taipei; TENMAK, T\"urkiye; STFC/UKRI, United Kingdom; DOE and NSF, United States of America.

Individual groups and members have received support from BCKDF, CANARIE, CRC and DRAC, Canada; CERN-CZ, FORTE and PRIMUS, Czech Republic; COST, ERC, ERDF, Horizon 2020, ICSC-NextGenerationEU and Marie Sk{\l}odowska-Curie Actions, European Union; Investissements d'Avenir Labex, Investissements d'Avenir Idex and ANR, France; DFG and AvH Foundation, Germany; Herakleitos, Thales and Aristeia programmes co-financed by EU-ESF and the Greek NSRF, Greece; BSF-NSF and MINERVA, Israel; Norwegian Financial Mechanism 2014-2021, Norway; NCN and NAWA, Poland; La Caixa Banking Foundation, CERCA Programme Generalitat de Catalunya and PROMETEO and GenT Programmes Generalitat Valenciana, Spain; G\"{o}ran Gustafssons Stiftelse, Sweden; The Royal Society and Leverhulme Trust, United Kingdom.

In addition, individual members wish to acknowledge support from CERN: European Organization for Nuclear Research (CERN PJAS); Chile: Agencia Nacional de Investigaci\'on y Desarrollo (FONDECYT 1190886, FONDECYT 1230812, FONDECYT 1230987); China: Chinese Ministry of Science and Technology (MOST-2023YFA1605700), National Natural Science Foundation of China (NSFC - 12175119, NSFC 12275265, NSFC-12075060); Czech Republic: Czech Science Foundation (GACR - 24-11373S), Ministry of Education Youth and Sports (FORTE CZ.02.01.01/00/22\_008/0004632), PRIMUS Research Programme (PRIMUS/21/SCI/017); EU: H2020 European Research Council (ERC - 101002463); European Union: European Research Council (ERC - 948254, ERC 101089007), Horizon 2020 Framework Programme (MUCCA - CHIST-ERA-19-XAI-00), European Union, Future Artificial Intelligence Research (FAIR-NextGenerationEU PE00000013), Italian Center for High Performance Computing, Big Data and Quantum Computing (ICSC, NextGenerationEU); France: Agence Nationale de la Recherche (ANR-20-CE31-0013, ANR-21-CE31-0013, ANR-21-CE31-0022, ANR-22-EDIR-0002), Investissements d'Avenir Labex (ANR-11-LABX-0012); Germany: Baden-Württemberg Stiftung (BW Stiftung-Postdoc Eliteprogramme), Deutsche Forschungsgemeinschaft (DFG - 469666862, DFG - CR 312/5-2); Italy: Istituto Nazionale di Fisica Nucleare (ICSC, NextGenerationEU), Ministero dell'Università e della Ricerca (PRIN - 20223N7F8K - PNRR M4.C2.1.1); Japan: Japan Society for the Promotion of Science (JSPS KAKENHI JP21H05085, JSPS KAKENHI JP22H01227, JSPS KAKENHI JP22H04944, JSPS KAKENHI JP22KK0227); Netherlands: Netherlands Organisation for Scientific Research (NWO Veni 2020 - VI.Veni.202.179); Norway: Research Council of Norway (RCN-314472); Poland: Ministry of Science and Higher Education (IDUB AGH, POB8, D4 no 9722), Polish National Agency for Academic Exchange (PPN/PPO/2020/1/00002/U/00001), Polish National Science Centre (NCN 2021/42/E/ST2/00350, NCN OPUS nr 2022/47/B/ST2/03059, NCN UMO-2019/34/E/ST2/00393, UMO-2020/37/B/ST2/01043, UMO-2021/40/C/ST2/00187, UMO-2022/47/O/ST2/00148, UMO-2023/49/B/ST2/04085); Slovenia: Slovenian Research Agency (ARIS grant J1-3010); Spain: Generalitat Valenciana (Artemisa, FEDER, IDIFEDER/2018/048), Ministry of Science and Innovation (MCIN \& NextGenEU PCI2022-135018-2, MICIN \& FEDER PID2021-125273NB, RYC2019-028510-I, RYC2020-030254-I, RYC2021-031273-I, RYC2022-038164-I), PROMETEO and GenT Programmes Generalitat Valenciana (CIDEGENT/2019/023, CIDEGENT/2019/027); Sweden: Swedish Research Council (Swedish Research Council 2023-04654, VR 2018-00482, VR 2022-03845, VR 2022-04683, VR 2023-03403, VR grant 2021-03651), Knut and Alice Wallenberg Foundation (KAW 2018.0157, KAW 2018.0458, KAW 2019.0447, KAW 2022.0358); Switzerland: Swiss National Science Foundation (SNSF - PCEFP2\_194658); United Kingdom: Leverhulme Trust (Leverhulme Trust RPG-2020-004), Royal Society (NIF-R1-231091); United States of America: U.S. Department of Energy (ECA DE-AC02-76SF00515), Neubauer Family Foundation.

%
%

% End of text imported from the .//acknowledgements/Acknowledgements.tex input file

%
%
%
%
%
%
%

%
\clearpage

\part*{Appendix}
\addcontentsline{toc}{part}{Appendix}

\begin{figure}[ht!]
\centering
\includegraphics[width=\linewidth]{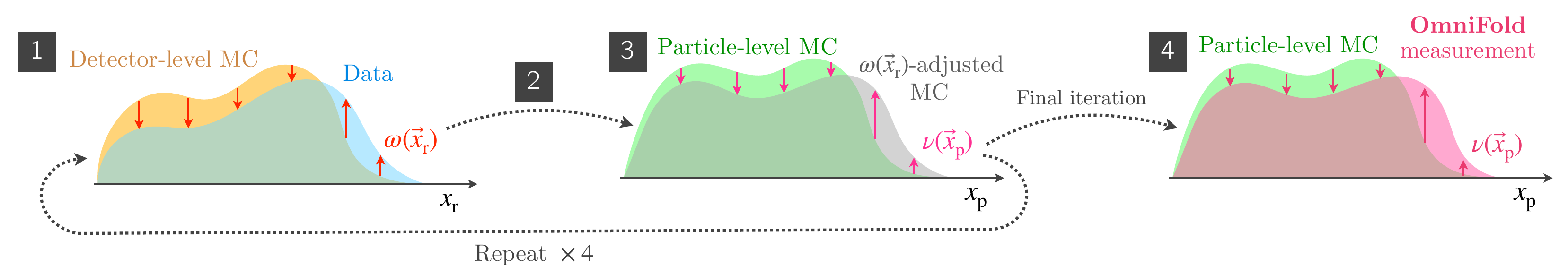}
\caption{Illustration of the \omnifold{} method. First, MC is corrected to match data at the detector level. Second, particle-level MC is adjusted by propagating the learned correction through the MC using a weighting function $\omega(\vec{x}_\mathrm{r})$. Third, a new correction $\nu(\vec{x}_\mathrm{p})$ is learned based on particle-level quantities only. Finally, $\nu(\vec{x}_\mathrm{p})$ is propagated through the MC back to the detector level achieving an improved agreement to data. The method proceeds iteratively four more times,
achieving a combined function $\nu(\vec{x}_\mathrm{p})$ that reweights the MC such that the event yields and kinematics match those observed in the data.
\label{fig:omnifold}
}
\end{figure}

The \omnifold{} method~\cite{Andreassen:2019cjw,Andreassen:2021zzk} is illustrated in Figure~\ref{fig:omnifold} and briefly reviewed here. The approach takes two event samples as input: (1) an MC sample containing events with both particle-level ($\vec{x}_\mathrm{p}$) and detector-level ($\vec{x}_\mathrm{r}$) information, and (2) the reconstructed data, where $\vec{x}$ is a list of the 24 event-level observables: $\vec{x}=(\ptll,\dots,\tausj{3})$.
The MC sample encodes information about the detector response such as energy and momentum resolution, particle and selection inefficiencies.
The method is based on a multidimensional reweighting: at each step, a smooth weighting function is obtained using the event observables $\vec{x}$.
First, it corrects reconstructed-level MC to match data with weights $\omega(\vec{x_\mathrm{r}})$.
Next, an improved MC prediction is achieved by propagating $\omega(\vec{x}_\mathrm{r})$ to the particle level $\vec{x}_\mathrm{p}$. Then, a new reweighting function $\nu(\vec{x}_\mathrm{p})$ is obtained by reweighting the particle-level prediction to the improved one from the previous step. It should be noted that $\nu(\vec{x}_\mathrm{p})$ itself does not rely on the detector level input, even if this was crucial in the derivation of it.
Event weights defined by $\nu(\vec{x}_\mathrm{p})$ is then propagated back through the MC to the detector level, resulting in an improved prediction to the data compared to the initial MC. The method is repeated iteratively; the updated reweighting function becomes a product of the previous one until a predefined number of iterations are performed, when the method stops.

The reweighting functions used in \omnifold{} are based on the output $f(\vec{x})$ of NNs trained with a weighted binary cross-entropy loss function:
\begin{align}
\label{eq:loss}
\mathcal{L}[f(\vec{x})] = -\sum_{i\in A} w_i\log(f(\vec{x}_i)) - \sum_{i\in B} w_i \log(1-f(\vec{x}_i))\,,
\end{align}
where the $w_i$ are event weights for datasets $A$ and $B$, with associated (joint) probability densities $p_A(\vec{x})$ and $p_B(\vec{x})$.
With this choice of loss function, the produced NN classifier $f(\vec{x})$ can be
used to define the quantity
$\omega(\vec{x}) \equiv f(\vec{x})/(1-f(\vec{x}))$, which is known to asymptotically approach the likelihood ratio $p_A(\vec{x})/p_B(\vec{x})$~\cite{Hastie:2001,Sugiyama:2012}, assuming $A$ and $B$ are normalized to unity ($\sum_A w_i=\sum_B w_i=1$). This quantity is used to perform the 24-dimensional shape reweighting in the first and third step of the \omnifold{} method. The correction in the second and fourth step is applied to each event $i$ in sample $B$ by updating the event weights by $w_i \mapsto w_i\,\omega(\vec{x}_i)$.

In the first reweighting step of \omnifold, Sample $A$ is data and Sample $B$ is detector-level MC simulation. Each event weight $w_i=1$ for data, while for MC simulation, $w_i$ is initialized as the nominal MC sample weights.
Both samples are normalized such that $\sum_i w_i =1$, and the MC event weights are updated for each iteration as described above. In the second reweighting step of \omnifold, Samples $A$ and $B$ both have the same events and kinematics $\vec{x}_\mathrm{p}$, but the weights for $A$ are taken as $\omega(\vec{x}_r)$ from the previous step. The reweighting then takes place using only particle-level quantities.

This analysis does not subtract the background as it is small ($<0.25\%$). A natural way to subtract backgrounds as part of the \omnifold{} procedure with nontrivial backgrounds would be to add negatively-weighted MC events to the $A$ dataset, such that in total, $A$ corresponds to data with background subtracted. ~\cite{Andreassen:2021zzk}.
The differential acceptance and efficiency are accounted for naturally by events satisfying one of the detector-level or particle-level event selections, but not both.
Events that do not carry an updated weight from the previous step are assigned the average weight in their region of phase space $x$ using Eq.~(\ref{eq:loss})~\cite{Andreassen:2021zzk}.

\printbibliography

\clearpage

\part*{Supplementary Material}
\addcontentsline{toc}{part}{Supplementary Material}

This supplement provides further details about how to use the unbinned \omnifold{} measurement that is presented in the Letter.
The datasets are publicly available via Refs.~\cite{public_multifold, zenodo} along with Jupyter notebooks that provide detailed examples of how to use the results,
including obtaining the measured cross sections, covariance matrices, and plots of differential cross sections and the uncertainty breakdown.

\section{Description of the datasets}
The published datasets are event samples formatted as Pandas DataFrames with a total size of about 5 GB.
Each event contains the 24 kinematic particle-level observables and a series of event weights.
All weights have femtobarn units such that summing them for a given kinematic region produces the corresponding fiducial cross section.
There is a nominal event weight (\texttt{weights\_nominal}) that provides the central value of the measurement, and a long list of weights that each correspond to the uncertainty variations.

The repository includes the following notebooks:
\begin{enumerate}
\item \texttt{ATLAS\_Zjets\_MultiFold\_Basics.ipynb}, which provides instructions for initial setup and an overview of the content and basic usage of the datasets;
\item \texttt{ATLAS\_Zjets\_MultiFold\_PseudoResults.ipynb}, which shows how to use the measurement of the pseudo-data and its target distribution that can be used for analysis validation; and
\item \texttt{ATLAS\_Zjets\_MultiFold\_Results.ipynb}, which shows how to use the measurement based on actual data, including how to reproduce the plots shown in the Letter.
\end{enumerate}

\section{Usage recommendations}
\label{sec:usage}
As detailed in the paper, the unbinned measurements were validated both in the full phase space ($\ptll>\SI{200}{\GeV}$) and in various kinematic sub-regions.
In these regions, a series of binned differential cross sections were studied, both for the 24 directly measured observables and additional observables derived as a function of the 24, such as $m_{jj}$, $\Delta y_{jj}$ and $\Delta R(\ell\ell,j_1)$.
When using these results, the following guidance is recommended:

\begin{description}

\item[1. Jet $p_\text{T}$ threshold]~\\When studying jet observables, a jet \pt{} threshold should be applied of at least \SI{5}{\GeV}. For example, if plotting $\Delta R(\ell\ell,j_1)$, which relies on the leading jet, a requirement should be placed on the leading jet \pt{}, and for $m_{jj}$, \pt{} requirements should be applied to both jets.
\item[2. Sufficient Monte Carlo statistical precision]~\\Each bin should contain a minimum of 5,000 effective events defined by $n_{\text{eff}} \equiv (\Sigma_iw_i)^2 / \Sigma _iw_i^2$.
This ensures that the bin has sufficient statistical precision, which is important for stabilizing the bin uncertainties.
\item[3. Sufficient data statistical precision]~\\Each bin should further have a data statistical uncertainty smaller than $15\%$, which ensures that the kinematic region probed has sufficient support by the data.
This criterion is most likely fulfilled when $n_{\text{eff}} > 5$,000.
\item[4. Validation using the pseudo-measurement]~\\Any analysis should always be checked using the pseudo-measurement against the target, to ensure a reliable result is obtained before drawing any conclusion based on the measurement with real data. Examples of such tests are performed in \texttt{ATLAS\_Zjets\_MultiFold\_PseudoResults.ipynb}.

The tests performed as part of the example code is a standard, binned $\chi^2$ compatibility test between the pseudo-measurement and the target.
The key step of this procedure is to construct the measurement covariance matrix that encodes the correlation between bins.
The $\chi^2$ and associated $p$-value are then obtained with a few lines of code that performs matrix operations.
The outcome of these tests will depend on the assumed correlation model as discussed in the next bullet.

\item[5. Alternative treatment of uncertainty correlation]~\\
The default recommended procedure is to treat each systematic uncertainty component as fully correlated across bins. However, this might not be ideal for two-point systematic uncertainties, for which the internal uncertainty correlation is not known. Certain regions of phase space have dominant uncertainty from the two unfolding systematic uncertainty sources, which both are two-point systematic uncertainties.
This can in particular be an issue when using \msj, where the uncertainty amplitude has a strong shape dependence, and a switch in sign in the middle of the distribution.
In cases like this, poor $p$-values may sometimes result from assuming bins are fully correlated.

In these cases, other correlation models may be used. As the true correlation model is not known, any correlation assumption must be tested on the pseudodata before it can be used to extract information from the actual data. If few bins are used, one can treat the bins as fully uncorrelated for this uncertainty component. For instance, if four bins are used, the systematic uncertainty can be split into four components (`nuisance parameters'). More sophisticated decorrelation schemes could be explored for which each uncertainty component has a smooth uncertainty amplitude spanning the full range where at any point (i.e.\ for any bin) this uncertainty adds in quadrature to match the total uncertainty of the two-point systematics.

\end{description}

\section{Extracting particle-level results}

The notebook titled \texttt{ATLAS\_Zjets\_MultiFold\_Results.ipynb} can be used to not only reproduce the results shown in the Letter, but also to create a variety of other distributions. The default binning choice used in the paper is implemented in \texttt{multifold\_util.py}. The first part of this notebook is dedicated to calculating the bin uncertainties for each observable and uncertainty source. The general strategy for calculating systematic uncertainties is displayed below:
\begin{lstlisting}[language=Python, caption=Python pseudocode for obtaining binning and the uncertainty of a given systematic uncertainty source.]
# pick or loop over observables ("var") added to dataframe ("df")
# set binning with array bins
for var in variable_list:
bins = np.array(bins_dict[var]) # bins_dict can be modified to change binning

## the central value of the bin (fiducial) cross section
nom, _  = np.histogram(df[var], bins=bins, weigts=df.weights_nominal)
## the corresponding cross section for a systematic variation.
## This applies to 22 out of the 29 uncertainty sources. "syst" below needs to be replaced with e.g. "muonCal"
syst, _ = np.histogram(df[var], bins=bins, weights=df.weights_syst)
## systematic fractional uncertainty amplitude of a given source ("nuisance parameter"). To be used for e.g. covariance matrix
syst_lambda = (syst-nom)/nom
# absolute, percentage uncertainty; for plotting
syst_uncert = 100*np.sqrt((syst-nom)**2)/nom
\end{lstlisting}

\begin{figure*}[h!]
\centering
\includegraphics[width=\linewidth]{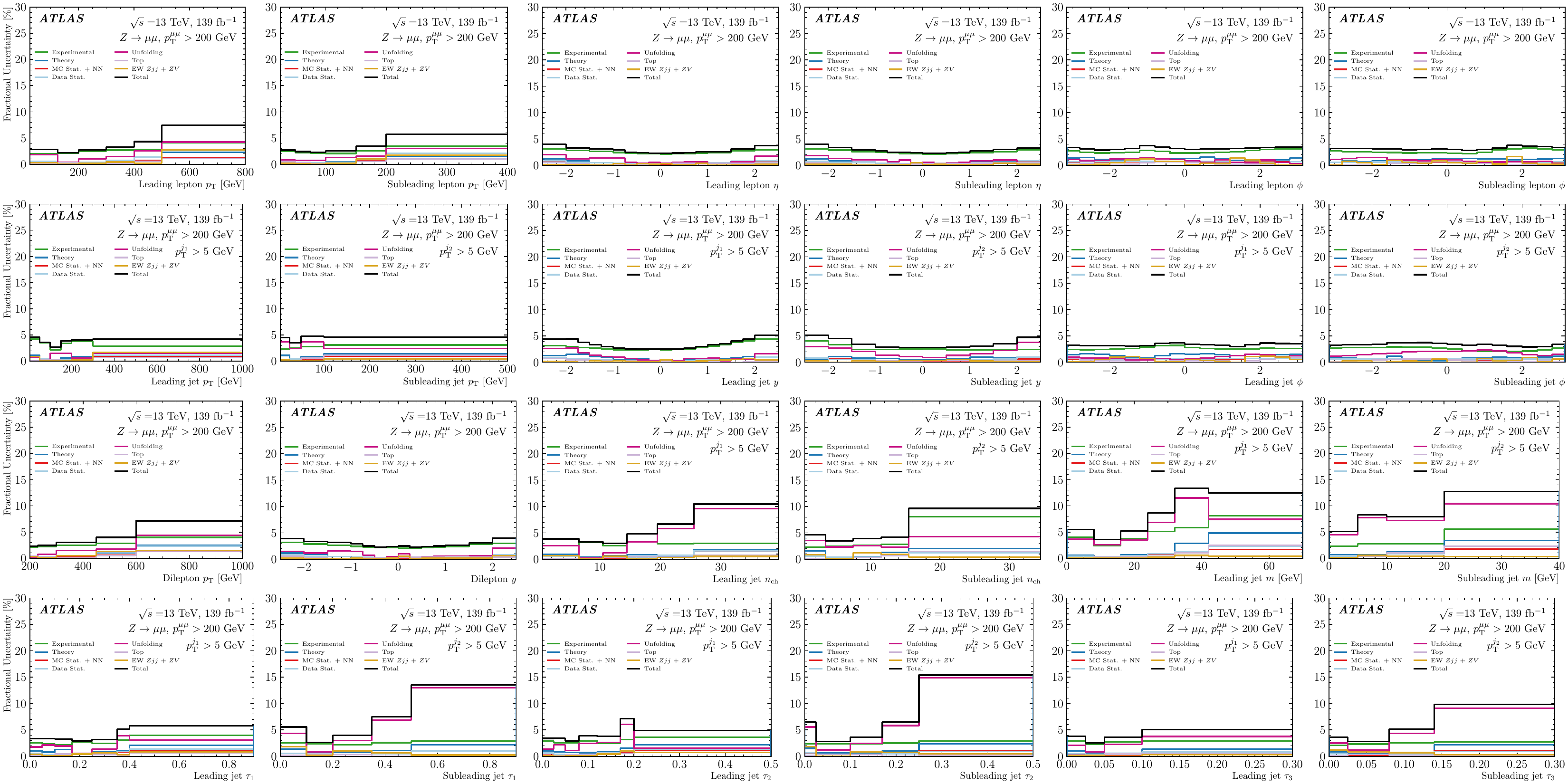}
\caption{The total uncertainty and its breakdown into various categories of the 24 directly measured $Z$+jet observables measured in this analysis. }
\label{fig:syst24}
\end{figure*}

The total uncertainty and its breakdown into main categories are shown for the differential cross sections of the 24 directly measured observables in Figure~\ref{fig:syst24}.
The default binning choice is used, however this can easily be modified.

\begin{figure*}[h!]
\centering
\includegraphics[width=\linewidth]{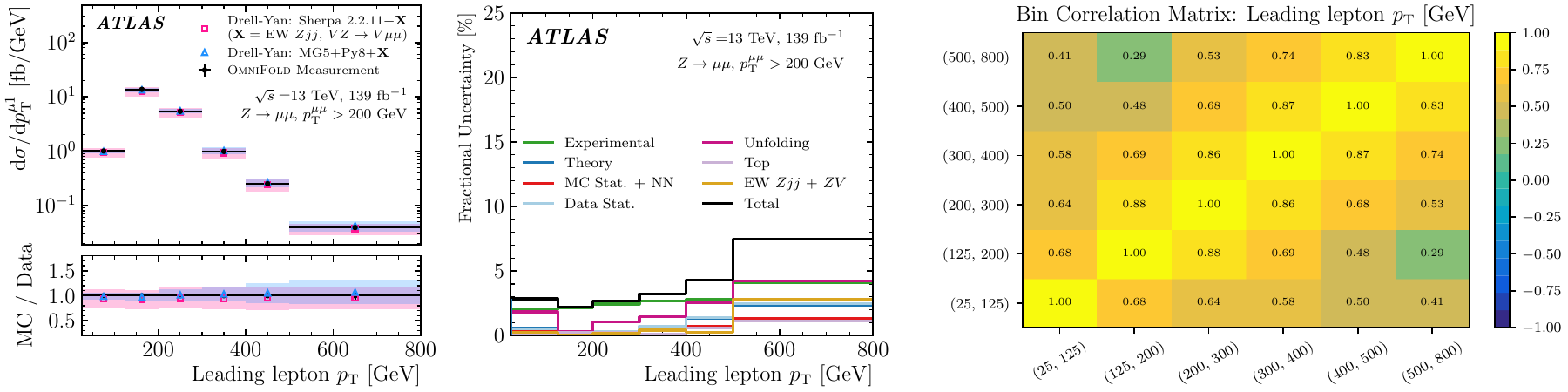}
\includegraphics[width=\linewidth]{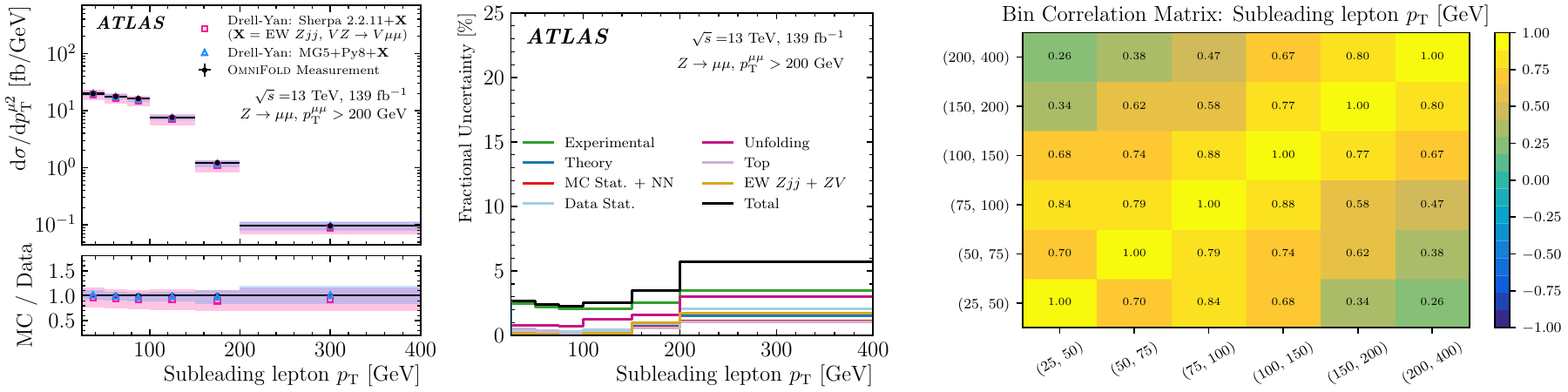}
\caption{Measured differential cross sections along with predictions from \MADGRAPH{} and \SHERPA[2.2.11] (left), the total measurement uncertainty and its breakdown into sources (middle), and the associated correlation matrix (right) for \ptlm{} (top) and \ptsm{} (bottom).}
\label{fig:corr}
\end{figure*}

Figure~\ref{fig:corr} shows the differential cross-section measurements, the uncertainty breakdown into various categories, and the associated correlation matrices for two example variables: \ptlm{} and \ptsm{}.

\begin{figure*}[p]
\centering
\includegraphics[width=\linewidth]{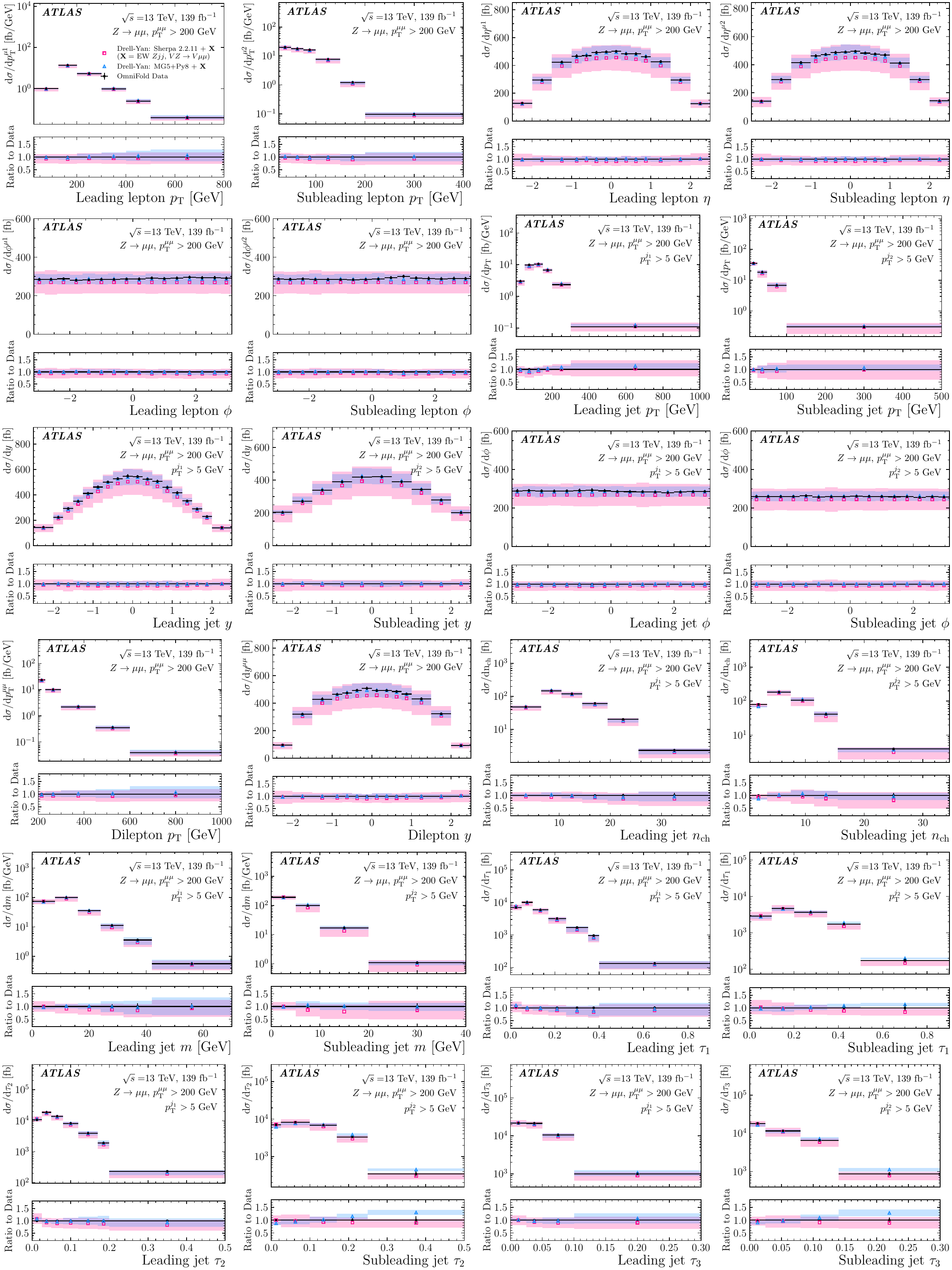} %
\caption{Measured differential cross sections of the 24 $Z$+jet observables compared with particle-level predictions.  For display purposes, binned (marginal) distributions are shown, though the measurement itself is unbinned and 24-dimensional.}
\label{fig:nom24}
\end{figure*}

Once the uncertainties are calculated, users can produce plots of differential cross-sections, such as those shown in Figure~\ref{fig:nom24} using code of the form:
\begin{lstlisting}[language=Python, caption=Python pseudocode for plotting a differential cross-section.]
import matplotlib.pyplot as plt

# pick or loop over observables ("var") added to dataframe ("df")
# set binning with array bins
bin_centers = 0.5 * (bins[1:] + bins[:-1])
bin_widths = np.array([bins[n+1] - bins[n] for n in range(len(bins)-1)])

multifold_density, _, _ = plt.hist(df[var], bins=bins, density=True, weights=df.weights_nominal, color="k", alpha=0)
multifold_density *= np.sum(df.weights_nominal)

# run cells in notebook to compute the uncertainties dictionary, based on functions like calculate_uncertainty
plt.errorbar(bin_centers, multifold_density, xerr=bin_widths/2, yerr=multifold_density*uncertainties[var+"_total"]/100, label="MultiFold")

\end{lstlisting}

The binning choice can easily be modified as illustrated below, however one need to ensure to follow the guidelines as outlined in Section~\ref{sec:usage}.
\begin{lstlisting}[language=Python, caption=Changing bins for a particular observable]
bins_dict = ibu_bins # defined in demo_functions.py
bins_dict["Ntracks_trackj1"] = [0.5, 2] # example of how to modify binning for a given variable
\end{lstlisting}

\begin{figure*}[h!]
\centering
\includegraphics[width=0.32\linewidth]{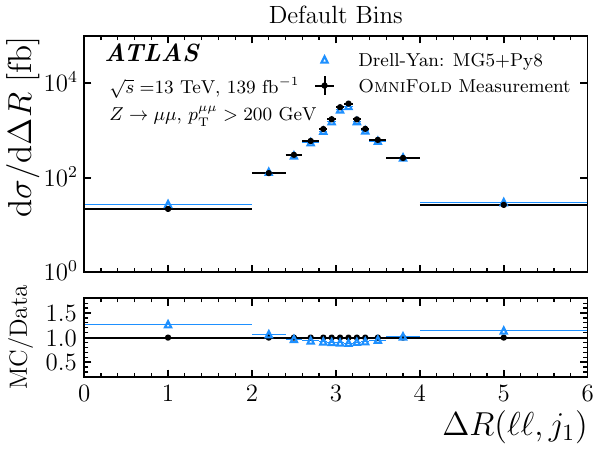}
\includegraphics[width=0.32\linewidth]{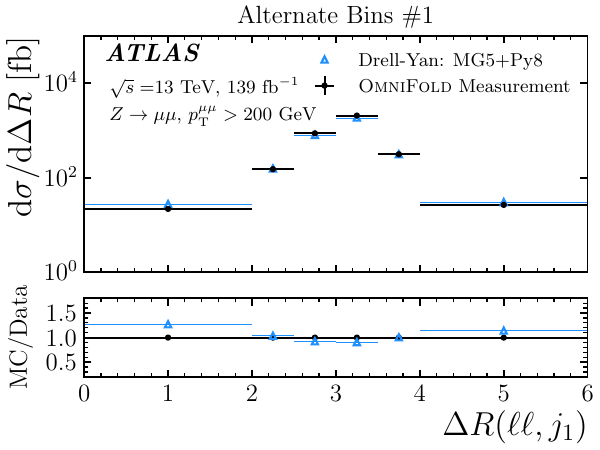}
\includegraphics[width=0.32\linewidth]{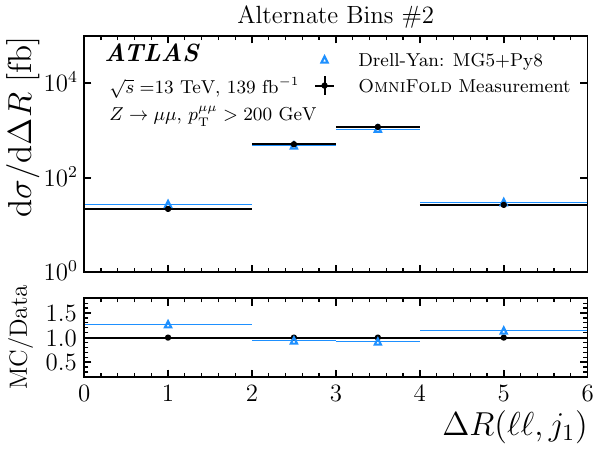}
\caption{Measured differential cross sections of $\Delta R(\ell\ell,j_1)$ using three different bin choices. MC predictions are only shown with statistical uncertainties.}
\label{fig:dr_bins}
\end{figure*}

Figure~\ref{fig:dr_bins} presents three distributions of the derived variable $\Delta R(\ell\ell,j_1)$ with different choices of bins.

\begin{figure*}[h!]
\centering
\includegraphics[width=0.7\linewidth]{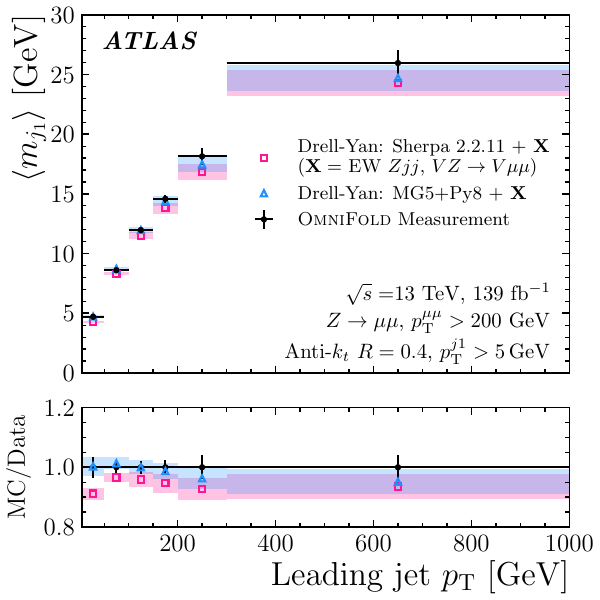}
\caption{The average measured leading jet $m$ as a function of jet \pt{} and the associated predictions from \MADGRAPH{} and \SHERPA[2.2.11].}
\label{fig:ave_pt}
\end{figure*}

These datasets can also be used to produce measurements of quantities that are not differential cross sections. For instance, Figure~\ref{fig:ave_pt} shows distributions of the average jet mass in bins of jet \pt{} for the leading track jet.

\clearpage
% ATLAS Collaboration author list
% Reference date of STDM-2020-17 is 2024-03-07
% Author list last updated on date 15-OCT-24
% Data extracted on 15-Oct-2024 for paper reference STDM-2020-17
% at 10:29am
 
\begin{flushleft}
\hypersetup{urlcolor=black}
{\Large The ATLAS Collaboration}

\bigskip

\AtlasOrcid[0000-0002-6665-4934]{G.~Aad}$^\textrm{\scriptsize 104}$,
\AtlasOrcid[0000-0001-7616-1554]{E.~Aakvaag}$^\textrm{\scriptsize 17}$,
\AtlasOrcid[0000-0002-5888-2734]{B.~Abbott}$^\textrm{\scriptsize 123}$,
\AtlasOrcid[0000-0002-0287-5869]{S.~Abdelhameed}$^\textrm{\scriptsize 119a}$,
\AtlasOrcid[0000-0002-1002-1652]{K.~Abeling}$^\textrm{\scriptsize 56}$,
\AtlasOrcid[0000-0001-5763-2760]{N.J.~Abicht}$^\textrm{\scriptsize 50}$,
\AtlasOrcid[0000-0002-8496-9294]{S.H.~Abidi}$^\textrm{\scriptsize 30}$,
\AtlasOrcid[0009-0003-6578-220X]{M.~Aboelela}$^\textrm{\scriptsize 45}$,
\AtlasOrcid[0000-0002-9987-2292]{A.~Aboulhorma}$^\textrm{\scriptsize 36e}$,
\AtlasOrcid[0000-0001-5329-6640]{H.~Abramowicz}$^\textrm{\scriptsize 155}$
% ...
\bigskip
\\

% ...
$^{17}$Department for Physics and Technology, University of Bergen, Bergen; Norway.\\
% ...
$^{30}$Physics Department, Brookhaven National Laboratory, Upton NY; United States of America.\\
% ...
$^{36}$$^{(a)}$Facult\'e des Sciences Ain Chock, Universit\'e Hassan II de Casablanca;$^{(b)}$Facult\'{e} des Sciences, Universit\'{e} Ibn-Tofail, K\'{e}nitra;$^{(c)}$Facult\'e des Sciences Semlalia, Universit\'e Cadi Ayyad, LPHEA-Marrakech;$^{(d)}$LPMR, Facult\'e des Sciences, Universit\'e Mohamed Premier, Oujda;$^{(e)}$Facult\'e des sciences, Universit\'e Mohammed V, Rabat;$^{(f)}$Institute of Applied Physics, Mohammed VI Polytechnic University, Ben Guerir; Morocco.\\
% ...
$^{45}$Physics Department, Southern Methodist University, Dallas TX; United States of America.\\
% ...
$^{50}$Fakult\"{a}t Physik , Technische Universit{\"a}t Dortmund, Dortmund; Germany.\\
% ...
$^{56}$II. Physikalisches Institut, Georg-August-Universit\"{a}t G\"ottingen, G\"ottingen; Germany.\\
% ...
$^{104}$CPPM, Aix-Marseille Universit\'e, CNRS/IN2P3, Marseille; France.\\
% ...
$^{119}$$^{(a)}$New York University Abu Dhabi, Abu Dhabi;$^{(b)}$United Arab Emirates University, Al Ain; United Arab Emirates.\\
% ...
$^{123}$Homer L. Dodge Department of Physics and Astronomy, University of Oklahoma, Norman OK; United States of America.\\
% ...
$^{155}$Raymond and Beverly Sackler School of Physics and Astronomy, Tel Aviv University, Tel Aviv; Israel.\\
% ...
$^{a}$ Also Affiliated with an institute covered by a cooperation agreement with CERN.\\
$^{b}$ Also at An-Najah National University, Nablus; Palestine.\\
$^{c}$ Also at Borough of Manhattan Community College, City University of New York, New York NY; United States of America.\\
$^{d}$ Also at Center for Interdisciplinary Research and Innovation (CIRI-AUTH), Thessaloniki; Greece.\\
$^{e}$ Also at Centro Studi e Ricerche Enrico Fermi; Italy.\\
$^{f}$ Also at CERN, Geneva; Switzerland.\\
% ...
$^{*}$ Deceased

\end{flushleft}

% Created with Glance <Atlas.Glance@cern.ch>

%

%
%
%
%

\end{document}